 \documentclass[journal,twoside]{IEEEtran}
\usepackage{ifpdf}
\usepackage{widetext}
\usepackage{cite}
\ifCLASSINFOpdf
   \usepackage[pdftex]{graphicx}
   \graphicspath{{fig/}}
\else
\fi
\usepackage[usenames,dvipsnames,svgnames,table]{xcolor}

\usepackage{amsmath}
\usepackage{color}
\usepackage{amsmath,amsthm, amsfonts,amssymb}

\usepackage{array,booktabs,makecell}
\usepackage{footnote}
\makesavenoteenv{tabular}

\usepackage{array}
\usepackage{multirow}
\usepackage{colortbl}
\definecolor{kugray5}{RGB}{224,224,224}
\usepackage{balance}
\usepackage{color}
\usepackage{mathtools,lipsum,cuted}
\usepackage{cuted}
\begin{document}
\title{General Scattering Characteristics of Resonant Core-Shell Spheres}


\author{\IEEEauthorblockN{Dimitrios C. Tzarouchis,~\IEEEmembership{Student Member,~IEEE}, 
and Ari Sihvola,~\IEEEmembership{Fellow,~IEEE}}
%\IEEEauthorblockA{\IEEEauthorrefmark{1}Department of Radio Science and Engineering, Aalto University, 02150, Espoo, Finland}
% <-this % stops an unwanted space
% created January 30, 2016;
\thanks{Manuscript date: \today.
This work is supported by the Aalto Energy Efficiency research program (EXPECTS project) and the ELEC Doctoral School Scholarship. 
The authors are with the Department of Electronics and Nanoengineering (formerly known as Radio Science and Engineering), Aalto University, 02150, Espoo, Finland. (dimitrios.tzarouchis@aalto.fi)}}

% \markboth{IEEE Transactions on Antennas and Propagation}%
% {Tzarouchis and Sihvola\MakeLowercase{\textit{et al.}}: Pad\'{e} approximants for Mie coefficients}



% for Transactions on Magnetics papers, we must declare the abstract and
% index terms PRIOR to the title within the \IEEEtitleabstractindextext
% IEEEtran command as these need to go into the title area created by
\IEEEtitleabstractindextext{%
\begin{abstract}
This article presents and discusses the general features and aspects regarding the electromagnetic scattering by a small core-shell plasmonic sphere. First, the thickness effects on the plasmonic resonances are presented in the electrostatic (Rayleigh) limit, utilizing the MacLaurin expansion of the Mie coefficients of hollow scatterers. Several aspects regarding the core effects are given, illustrating the enabling mechanisms and peculiarities of its resonant scattering response on it electrostatic limit. {The electrodynamic aspects of the scattering process are revealed through the newly introduced Pad\'e expansion of the Mie coefficients. Additionally we expose how the core material affects the dynamic mechanisms, such as the dynamic depolarization and radiative damping.} The described method can be expanded for other type of resonances and canonical shapes, while the general characteristics presented here are expected to stimulate further studies regarding the functionalities of the core-shell scatterers. 
\end{abstract}

}

\maketitle

\IEEEdisplaynontitleabstractindextext

% \fi
\IEEEpeerreviewmaketitle

\section{Introduction}
\IEEEPARstart{T}{he} quest of controlling the electromagnetic radiation pushes modern science and engineering into uncharted territories, spanning from few millimeters up to hundred nanometers of wavelength. These novel technological achievements exploit concepts and phenomena occurring either by the single scatterer functionalities~\cite{Lannebere2015a,Alu2009a} or emerge as collective effects induced by the building block's properties~\cite{PhysRevX.5.031005,Jahani2016,Mazor2017}. In every case the study of the fundamental properties of individual scatterers plays a crucial role to the overall design process. 

One of the most studied canonical problems facilitating this purpose is the study of the electromagnetic scattering by a spherical scatterer; a long-studied benchmarking platform providing insights regarding the nature of the scattering phenomena~\cite{Schebarchov2013,Shore2015,Fan2014,Liberal2014,Lukyanchuk2010}. In this article, we cast light to the special case of a small core-shell spherical scatterer, describing its scattering attributions and resonant peculiarities.

\begin{figure}[!]
\centering
 \includegraphics[width=0.45\textwidth]{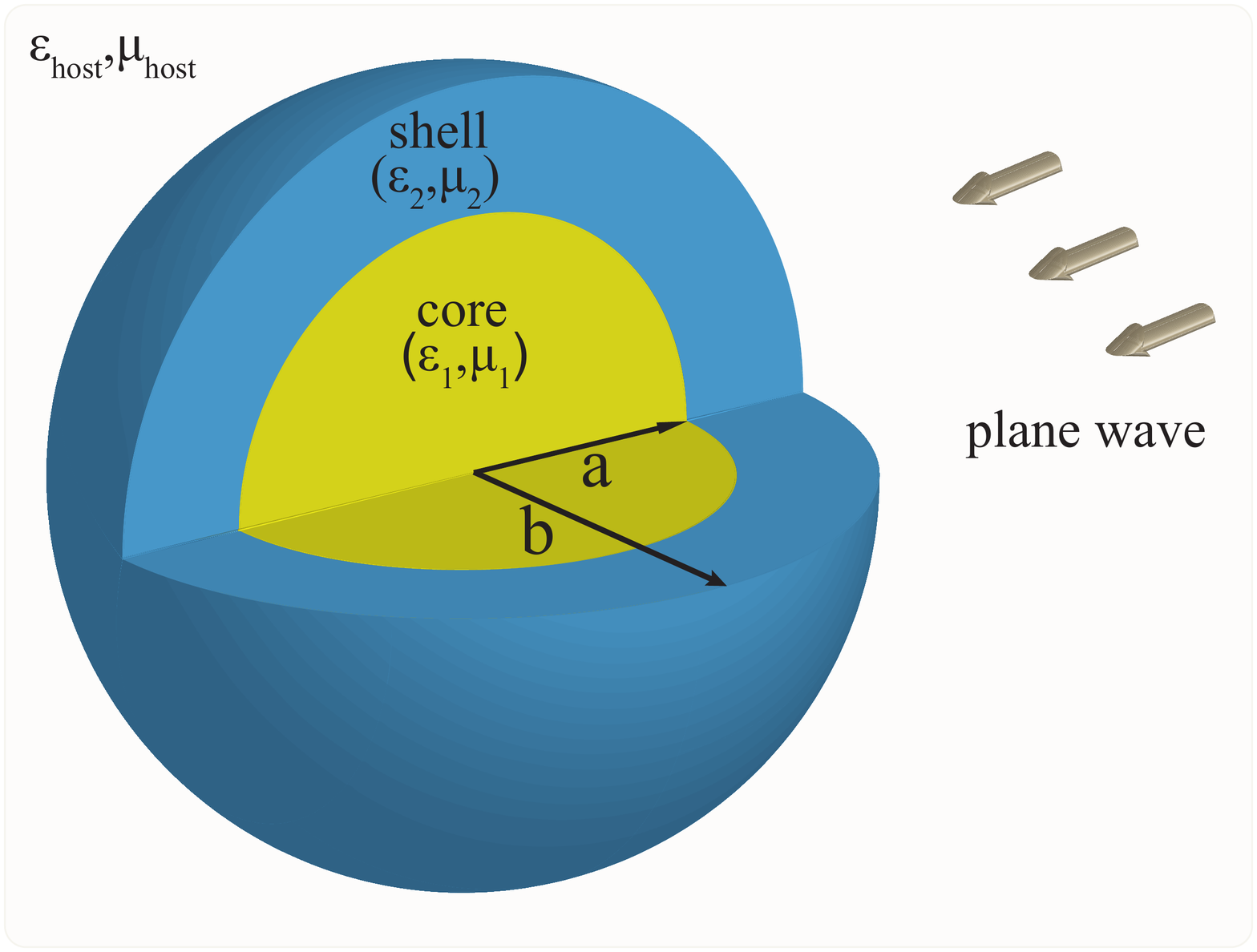}~
 \caption{Problem setup: a core-shell sphere immersed in a host medium with internal and external radius $a$ and $b$, under plane wave illumination. Region 1 corresponds to the core while region 2 to the shell region.}
 \label{fig:structure}
\end{figure}

Let us begin by assuming a monochromatic, linearly polarized plane wave\footnote{the time-harmonic convention $e^{-i\omega t}$ is used} impinging on a spherical, composite obstacle, as can be seen in Fig.~\ref{fig:structure}. Briefly, the field can be decomposed as a sum of spherical TE and TM field harmonics in every domain (also known as H- and E-waves)~\cite{stratton2007electromagnetic}. By applying the appropriate boundary conditions and solving the formulated problem, a set of field components occurs, described by the Mie coefficients~\cite{stratton2007electromagnetic}. These coefficients quantify the amplitude of each harmonic and their behavior. Our efforts will be mainly concentrated in studying the characteristics of the external scattering coefficients, denoted as $a_n$ and $b_n$. 

The $a_n$ coefficients account for the electric type multipole contributions, while the $b_n$ for the magnetic type. Both are expressed as a ratio of spherical Bessel and Riccati--Bessel functions, found readily in several classical textbooks~\cite{kerker1971scattering,bohren2008absorption}, and articles~\cite{Shore2015}. For magnetically inert case ($\mu_\text{host}=\mu_1=\mu_2=1$) the coefficients are affected by the (relative) permittivity, $\varepsilon_{\text{host}}$, the shell and core material $\varepsilon_{2}$ and $\varepsilon_{1}$, the core-to-shell radius ratio $\eta=\frac{a}{b}$, and the internal and external size parameters $x=ka$ and $y=kb$, with $k$ being the host medium wavenumber (see Fig.~\ref{fig:structure}).

Both $a_n$ and $b_n$ exhibit maximum and minimum values for a variety of material and morphological cases~\cite{bohren2008absorption}. A special type of resonances occurring for negative permittivity values of the shell are called localized surface plasmonic resonances (LSPR or plasmonic resonances). LSPR occur naturally for metals in the visible--near-infrared regime due to the collective oscillations of the free electrons~\cite{kreibig1995optical}. These electric charge oscillations are characterized as electric multipole resonances, visible in the spectrum of the $a_n$ coefficients. Our main focus will be to study the qualitative behavior of these resonances by studying the electric Mie coefficients. 

{
This work will be presented in the following way. First, a simple Taylor expansion of the $a_n$ coefficients will be given in Section~\ref{sec:static}, revealing the electrostatic aspects of the enabled resonances. In this classical perspective we present two limiting cases, for thin and thick shells, introducing their effects in the scattering process. In Section~\ref{sec:core} we expand the discussion on how the core material affects the electrostatic response. Section~\ref{sec:Dynamic} presents the newly introduced Mie-Pad\'e expansion for extracting intuitive results on the impact of the size-dependent dynamical mechanisms. 
}

{
The depolarization and radiative reaction effects will be presented, giving emphasis to the core material effects. The methodology described in the aforementioned sections is general and can be readily used for studying the behavior of other type of resonances, such as all-dielectric resonances~\cite{Kuznetsov2016}, or other type of canonical shapes, e.g., core-shell cylinders.}

\section{Resonant properties of core-shell plasmonic structures}\label{sec:static}

\begin{figure}
\centering
 \includegraphics[width=0.5\textwidth]{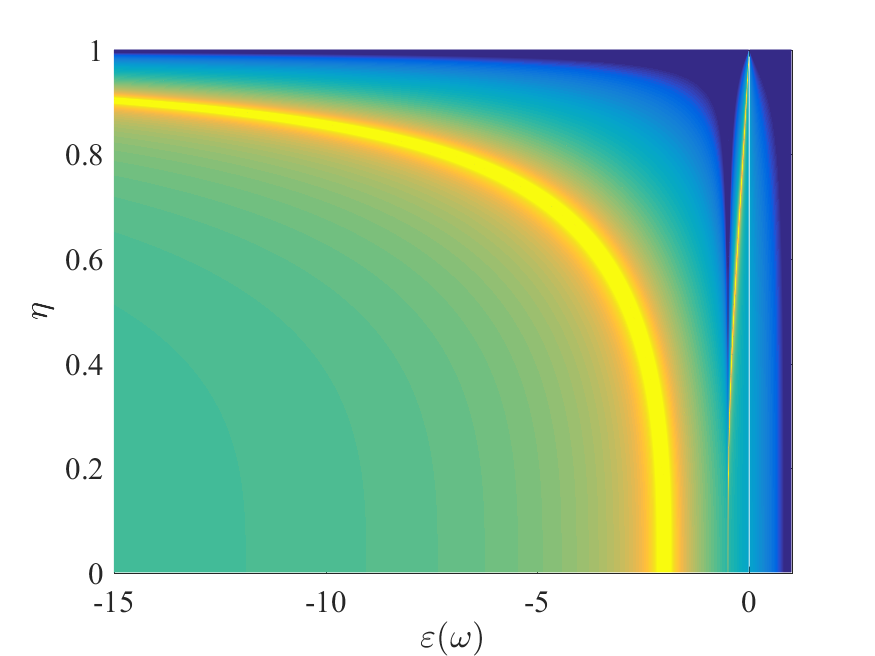}
 \caption{The extinction efficiency spectrum as a function of the permittivity of the shell (lossless) and the radius ratio $\eta$ for a small, hollow ($\varepsilon_\text{core}=1$) sphere ($y=0.01$). For the plasmonic case ($b_n\approx0$) the extinction efficiency is $Q_{\text{ext}}\propto\sum_n^\infty \Re\{a_n\}$, with $n=1,2,...$ . The bright lines correspond to the symmetric (left line) and antisymmetric (right line) dipole plasmonic resonances, while higher order multipoles are not visible. The color scale is logarithmic and is omitted, normalized for better visualization of the resonances.}
 \label{fig:sweep}
\end{figure}

To begin with, the MacLaurin (Taylor at $y=0$) series expansion of the first electric Mie coefficient ($a_1$) with respect to the size parameter $y=kb$ reads
\begin{equation} \label{eq:dpolar}
\begin{split}
&a_1^T=-i\frac{2}{3} \times \\
&\frac{(\varepsilon_2-1)(2\varepsilon_2+\varepsilon_1)-\eta^3(\varepsilon_2-\varepsilon_1)(2\varepsilon_2+1)}{(\varepsilon_2+2)(2\varepsilon_2+\varepsilon_1)-2\eta^3(\varepsilon_2-\varepsilon_1)(\varepsilon_2-1)}y^3+\mathcal{O}\left(y^5\right)
\end{split}
\end{equation}
In Fig.~\ref{fig:sweep} we can see the extinction efficiency
{
($Q_\text{ext}=\frac{2}{y^2}\sum^\infty_{n=1}\Re\{a_n+b_n\}$ where $a_n$ and $b_n$ are the electric and magnetic multipoles, respectively-see \cite{bohren2008absorption})} spectrum for the case of a small ($y=0.01$) hollow ($\varepsilon_1=1$) core-shell sphere as a function of the shell permittivity and the radius ratio $\eta$, where at least two resonances are visible.

It is clear that the Mie coefficient, in its Rayleigh limit (Eq.~(\ref{eq:dpolar})), is proportional to the volume-normalized static dipole polarizability, i.e., $a_{\text{Mie}}=-i\frac{2}{9}y^3\alpha_{\text{static}}$ with $\alpha_\text{static}=3\frac{(\varepsilon_2-1)(2\varepsilon_2+\varepsilon_1)-\eta^3(\varepsilon_2-\varepsilon_1)(2\varepsilon_2+1)}{(\varepsilon_2+2)(2\varepsilon_2+\varepsilon_1)-2\eta^3(\varepsilon_2-\varepsilon_1)(\varepsilon_2-1)}$~\cite{sihvola1999electromagnetic}, indicating that the full-electrodynamic Mie scattering model collapses to the electrostatic scattering description (Rayleigh) for vanishingly small spheres. Therefore, many physically intuitive observations are accessible through the analysis of the electrostatic model. From a mathematical perspective, Eq.~(\ref{eq:dpolar}) is nothing but a fractional function exhibiting zeros and poles. Physically, this function describes a system whose critical points correspond to scattering minima and maxima. The objective of this study is to extract information about the positions and the peculiarities of these points. 

% \subsection{Dipole effects of hollow spheres}
We will start by examining the simplest case of a hollow ($\varepsilon_1=1$), dielectric shell. The expanded Mie coefficient reads
\begin{equation}\label{eq:static_hollow}
 a_1^T\approx i\frac{2}{3}\frac{(\varepsilon_2-1)(2\varepsilon_2+1)(\eta^3-1)}{(\varepsilon_2+2)(2\varepsilon_2+1)-2\eta^3(\varepsilon_2-1)^2}y^3+\mathcal{O}\left(y^5\right)
\end{equation}
with pole condition (denominator zeros) described by

\begin{equation}\label{eq:static}
 \varepsilon^{\pm}=\frac{5+4\eta^3\mp3\sqrt{1+8\eta^3}}{4\left(\eta^3-1\right)}
\end{equation}
The above equation expresses the necessary condition for a pole to occur, namely the first electric dipole-like plasmonic resonances of a nanoshell~\cite{Averitt1999}. The notation $\varepsilon^{\pm}$ follows the plasmon hybridization model~\cite{Prodan2003}: $\varepsilon^-$ gives the permittivity value for the symmetric (bonding) and $\varepsilon^+$ the antisymmetric (antibonding) plasmonic resonance, respectively. Symmetric is the resonance with (symmetric) surface charge distribution between the inner and outer layer, with almost constant electric field in the inner regions, mainly residing in the outer surfaces. This type of resonance exhibits a plasmonic resonance at lower permittivity (energies) than the antisymmetric case, where it is mostly confined at the core region and the inner surfaces~\cite{Prodan2003}.

% \begin{figure*}[h]
% \centering
%  \includegraphics[width=1\textwidth]{new/fields_all.pdf}
%  \caption{Electric and magnetic field distribution (amplitude) of the symmetric (top) and antisymmetric (bottom) plasmonic resonances of a very small ($y=0.01$) core-shell sphere occurring for $\eta=0.1$ (left) and $\eta=0.9$ (right) radius ratio, respectively. The inset figures depict the scattering spectrum (extinction efficiency) for (a) $\eta=0.1$ and (b) for $\eta=0.9$. The center pictures illustrate the field distribution (electric and magnetic) of the plasmonic resonance for a solid sphere for comparison. An $E_x$ polarized, $k_z$ propagated plane wave is assumed.}
%  \label{fig:fields}
% \end{figure*}

Equation~(\ref{eq:static}) reveals that the required permittivity value is a function of the radius ratio. An even more intuitive condition can be extracted by expanding further Eq.~(\ref{eq:static}) and examining its behavior for two cases; for thick and thin shells. For the case of a thick shell, i.e.,  $\eta\rightarrow0$, we obtain two expressions, viz., 

\begin{equation}\label{eq:static_thick1}
 \varepsilon^+=-\frac{1}{2}+\frac{3}{2}\eta^3+\mathcal{O}\left( \eta^6 \right)
\end{equation}

\begin{equation}\label{eq:static_thick2}
 \varepsilon^-=-2-6\eta^3+\mathcal{O}\left( \eta^9 \right)
\end{equation}

These expressions describe a volume-dependent ($\eta^3$) behavior for the thick-shell resonances. {Assuming a thick shell expansion of the Eq.~(\ref{eq:static_hollow}) we obtain the following expression
 \begin{equation}
  a_1^T\approx-i\frac{2y^3}{3}+i\frac{2y^3}{\varepsilon_2+2}\left(1+\eta^3-\frac{6\eta^3}{\varepsilon_2+2}\right)+i\frac{2y^3\eta^3}{2\varepsilon_2+1}
 \end{equation}
 One notices that the first term is affected only by the size parameter, the second term represents the symmetric resonance $\varepsilon_2=-2$ condition for which a part is independent of the thickness $\eta$, and the third antisymmetric $\varepsilon_2=-\frac{1}{2}$ term with a thickness dependency. It is easy to see that for the limiting case $\eta\rightarrow0$ the antisymmetric term vanishes~\cite{sihvola2006character}.
}

The same expansion can be used for very thin shells ($\eta\rightarrow1$). For simplicity we define a new parameter the complementary ratio, defined as $\eta_c=1-\eta \in (0,1)$; vanishingly thin shells are described naturally for $\eta_c\rightarrow0$. We are now ready to express the pole condition of Eq.~(\ref{eq:static}) according to the complementary ratio as 
\begin{equation}\label{eq:static_thin1}
 \varepsilon^+=-\frac{2}{3}\eta_c+\mathcal{O}\left( \eta_c^2 \right)
\end{equation}

\begin{equation}\label{eq:static_thin2}
 \varepsilon^-=\frac{1}{2}-\frac{3}{2\eta_c}-\frac{1}{3}\eta_c+\mathcal{O}\left( \eta_c^2 \right)
\end{equation}
For $\eta_c\rightarrow0^+$ these resonant values approach $0$ and $-\infty$ respectively, which is an expected result. This linear behavior with respect to $\eta_c$ is qualitatively different from the thick-shell case where a volume dependence is observed.

% \subsection{General pole distribution for the plasmonic resonances: $n$-polarizabilities}\label{subseq:npol}

The expansion stratagem demonstrated above can be also used for higher order multipole resonances. Since the small-size expansions of the Mie coefficients boil down to the electrostatic polarizabilities, we generalize the above results following the static multipolarizabilities (or $n$-polarizabilities) extracted from e.g.,~\cite{Sihvola1988}. {A rather physical connection between the electrostatic limit and the plasmon hybridization is exposed in Appendix~\ref{sec:AppA}, especially for the thin- and thick-core limiting cases.}

As a rule, the $a_n$ Mie coefficients in the static limit can be written as functions of the pole order $n$ through the following expression

\begin{widetext}
\begin{equation}
 a_n= c_n \frac{((n+1) \varepsilon_2+n \varepsilon_1) (\varepsilon_2-1) -\eta ^{2 n+1} ((n+1) \varepsilon_2-n) (\varepsilon_2-\varepsilon_1)}{(n \varepsilon_2+n+1) ((n+1) \varepsilon_2+n \varepsilon_1)-n(n+1)\eta ^{2 n+1} (\varepsilon_2-1)  (\varepsilon_2-\varepsilon_1)}
\end{equation}
\end{widetext}

where $$c_n=-i\frac{n}{\left(2n-1\right)!!\left(2n+1\right)!!}y^{2n+1}$$ for $n=1,2,...$ and the higher order terms ($\mathcal{O}\left(y^{2n+3}\right)$) are truncated. From this relation, the thick-shells ($\eta\rightarrow0$) exhibit the following distribution (hollow core, $\varepsilon_1=1$)

\begin{equation}\label{eq:static2}
 \varepsilon^+=-\frac{n}{n+1}+\frac{n(2n+1)}{n+1}\eta^{2n+1}
\end{equation}
\begin{equation}\label{eq:static1}
 \varepsilon^-=-\frac{n+1}{n}-\frac{(n+1)(2n+1)}{n}\eta^{2n+1}
\end{equation}

The first term of Eq. (\ref{eq:static1}) corresponds to the known plasmonic distribution of a solid sphere, while the first term of Eq. (\ref{eq:static2}) follows the static resonances of the complementary problem, i.e., a hollow spherical cavity. Both resonances follow an $\eta^{2n+1}$ distribution with respect to the thickness. For large values of $n\rightarrow\infty$ both resonances  converge at $\varepsilon=-1$, implying a tendency of the higher order multipoles to accumulate in this value. However, higher multipoles exhibit very sharp linewidths, proportional to $y^{2n+1}$, and their effects are difficult to visualize~\cite{Tzarouchis2016c}.

The thin-shell analysis ($\eta_c\rightarrow0$) exposes a linear (and inverse linear) distribution for all multipoles, viz., 
\begin{equation}
 \varepsilon^+=-\frac{n\left(n+1\right)}{2n+1}\eta_c
\end{equation}

\begin{equation}
 \varepsilon^-=\frac{1}{n+1}-\frac{2n+1}{n\left(n+1\right)\eta_c}-\frac{n^2+n+1}{3\left(2n+1\right)}\eta_c
\end{equation}
suggesting a different character with that of the thick-shell case. This result can be readily used for the thick versus thin shell designing process. Note that the symmetric resonances are more sensitive ($\propto\frac{1}{\eta_c}$) to small thickness changes than the antisymmetric ones ($\propto\eta_c$). 

\section{Core material effects}\label{sec:core}

So far the radius ratio dependencies assuming a hollow shell were examined, i.e., with no material contrast between the host and core medium. In this section the dependencies caused by different core material contrasts will be discussed by allowing a general material description for the core (region 1), hereinafter denoted as $\varepsilon_1$. 

\begin{figure*}[h]
\centering
 \includegraphics[width=1\textwidth]{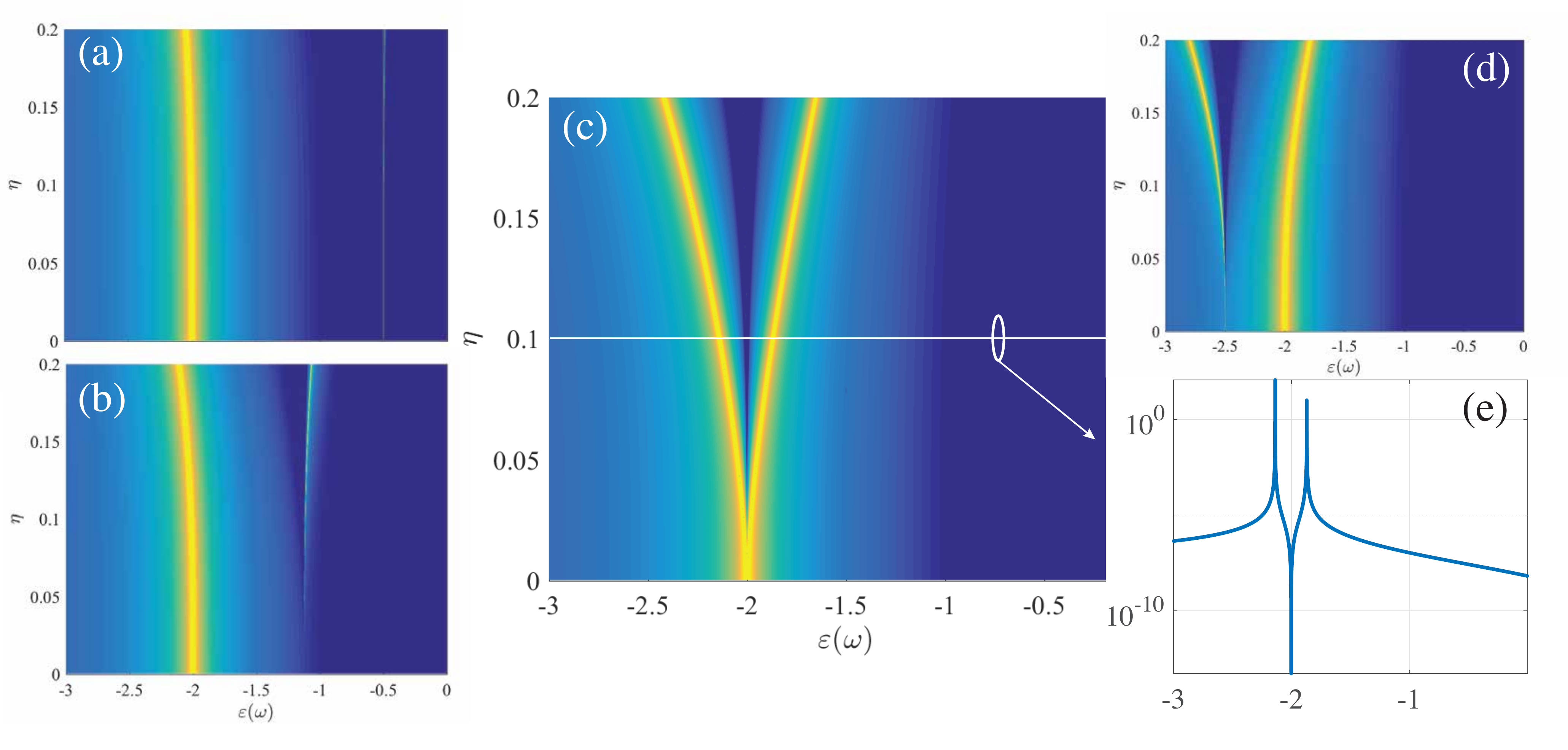}
 \caption{The scattering efficiency spectra for the case of (a) $\varepsilon_1=1$, (b) $\varepsilon_1=2.25$, (c) $\varepsilon_1=4$, and (d) $\varepsilon_1=5$. Note the behavior of the antisymmetric resonance, which for increasing core values exhibits a strong redshift. The extinction efficiency is in dB and the color range has been modified accordingly for depicting the resonant phenomena. Inset figure (e) depicts the extinction efficiency (in dB) for $\eta=0.1$, where one minimum and two maximum values are visible, exhibiting a Fano-like interference lineshape.}
 \label{fig:core}
\end{figure*}

For thick shells we distinguish two particular cases appearing on Eq.~(\ref{eq:dpolar}). This branching occurs for the core value $\varepsilon_1=4$, characterized as a critical value where both resonances degenerate to a single resonance for minuscule core-shell ratios, as in Fig.~\ref{fig:core} (c). Note that for the lossless case the scattering spectrum forms a Fano-like resonant profile with a scattering minimum in between both symmetric and antisymmetric resonant maxima (Fig.~\ref{fig:core} (e)). This can be also seen as a manifestation of the scattering equivalent of Foster's reactance theorem~\cite{Foster1924,Monticone2013a}. Furthermore, at this core value the two resonances deviate rapidly (Fig.~\ref{fig:core} (c)) for increasing radius ratio.  

{
It is interesting to notice that a general rule describing the core permittivity branching can be found, viz., 
\begin{equation}\label{eq:br}
\varepsilon^{\eta}_\text{br}=\left(\frac{n+1}{n}\right)^2 
\end{equation}
for every $n$-multipole. Here the superscript denotes the thick-shell case $\eta\rightarrow0$, implying that a different branching condition holds for the thin-shell case. This branching conditions reveals that even a small core shell might have a large impact on the position of the main resonances. Although the necessary core permittivity values can be empirically found by any Mie code implementation, condition of Eq.~(\ref{eq:br}) gives a rather straightforward rule regarding these degeneracy points.}

Going into the details, we obtain the two distinctive cases derived from Eq.~(\ref{eq:dpolar}), viz., 
\begin{alignat}{2}
\begin{rcases}
 \varepsilon^+=-\frac{\varepsilon_1}{2}-\frac{3}{2}\varepsilon_1\frac{\varepsilon_1+2}{\varepsilon_1-4}\eta^3
 \\
  \varepsilon^-=-2+6\frac{\varepsilon_1+2}{\varepsilon_1-4}\eta^3
\end{rcases}
~\text{for }~\varepsilon_1<4
 \end{alignat}
and
\begin{alignat}{2}
\begin{rcases}
 \varepsilon^+=-2+6\frac{\varepsilon_1+2}{\varepsilon_1-4}\eta^3
 \\
  \varepsilon^-=-\frac{\varepsilon_1}{2}-\frac{3}{2}\varepsilon_1\frac{\varepsilon_1+2}{\varepsilon_1-4}\eta^3
\end{rcases}
~\text{for }~\varepsilon_1>4
 \end{alignat}
Both symmetric and antisymmetric resonances mutually flip their attribution beyond the critical value. For the branching value both resonances exhibit a similar behavior, i.e,
\begin{equation}
 \varepsilon^{\pm}=-2\mp3\sqrt{2}\eta^{3/2}-\frac{9}{2}\eta^3
\end{equation}
where for $\eta\rightarrow0$ the two resonances degenerate to the known plasmonic resonant condition of a solid sphere, explaining the origin of this peculiarity. However, an interesting phenomenon occurs: the width of the antisymmetric resonance widens progressively as the core permittivity increases up to the value of $4$, as can be seen in Fig.~\ref{fig:core} (a), (b), and (c).

{For the thin-layer sphere a similar branching occurs at $\varepsilon_1=-2$, resulting in the following resonant conditions
\begin{alignat}{2}\label{eq:test}
\begin{rcases}
 \varepsilon^+=-2\frac{\varepsilon_1}{\varepsilon_1+2}\eta_c
 \\
  \varepsilon^-=\frac{\varepsilon_1}{2}-\frac{\varepsilon_1+2}{2\eta_c}
\end{rcases}
~\text{for }~\varepsilon_1>-2
 \end{alignat}
while for the $\varepsilon_1<-2$ the behavior changes mutually, similarly to the thick-shell. In this case the branching condition follows the rule 
\begin{equation}
\varepsilon^{\eta_c}_\text{br}=-\frac{n+1}{n} 
\end{equation}
which is exactly the position of the plasmonic $n$-multipole resonances of a solid sphere~\cite{sihvola2006character}. In the case of a core material with $\varepsilon_1=-2$, the core-shell sphere degenerates to a solid one. The main symmetric resonance is present while the antisymmetric one vanishes. Small deviations of this value cause the development of antisymmetric resonances, described in Eq.~(\ref{eq:test}).}

{
Another point of interest is how the core material affects the distribution of the scattering minima  (system zeros). For the case of a thick shell the zero condition is   $\varepsilon_\text{zero}=-\varepsilon_1\frac{n}{n+1}$ (with a trivial zero at $\varepsilon_\text{zero}=1$), while for the $\eta_c\rightarrow0$ case there are two zeros observed, i.e., for $\varepsilon_\text{zero}=\frac{\varepsilon_1}{n+1}-\frac{\varepsilon_1-1}{(n+1)\eta_c}$ and $\varepsilon_\text{zero}=\frac{n \varepsilon_1}{\varepsilon_1-1}\eta_c$, respectively.}

\section{Dynamic effects: Pad\'e approximants of Mie coefficients}\label{sec:Dynamic}
{
Sections~\ref{sec:static} and~\ref{sec:core} demonstrated the electrostatic properties of the plasmonic resonances, extended for accounting the core effects. These results are mainly based on the Taylor series expansions of the Mie coefficients, i.e., merely electrostatic expansions that catch several phenomena in scattering but fail to predict all size-induced dynamic effects. In this section we investigate how these dynamic effects contribute to the overall scattering response of the composite core-shell sphere.
}

{
The discussion regarding the size-induced dynamic effects has a long history, covered by classical textbooks and review papers, e.g., in~\cite{jackson1975electrodynamics,deVries1998}. These dynamic effects are implemented as corrections to the originally obtained electrostatic model, see for instance~\cite{Sipe1974,Meier1983,deVries1998,LeRu2013}. Briefly, there are two types of dynamical effects present in the small-size regime: the dynamic depolarization effect~\cite{Meier1983,deVries1998} and the radiation damping (reaction)~\cite{jackson1975electrodynamics,Sipe1974}. The first one accounts for the fact that within the small size limit the incident wave exhibits a constant phase across the scatterer and hence the field can be considered uniform. Apparently this does not hold for increasing sizes, when small changes on the polarization are introduced, and hence it is named as depolarization correction. The second type of correction is imposed by conservation of energy, i.e., the resonance of a lossless passive system cannot exhibit infinite amplitude values. This energy-conservation requirement is satisfied by inserting a term of $y^3$ order, independent of the material and geometrical characteristics of the scatterer~\cite{Kelly2003,Schebarchov2013,Tretyakov2014}.}

{For the case of a spherical scatterer, these corrections can be extracted in simple and straightforward manner by expanding the Mie coefficients utilizing the Pad\'e approximants. This idea has been recently introduced for the study of the dynamic scattering effects of small spheres~\cite{Tzarouchis2016c,Tzarouchis2017}. The Pad\'e approximants are a special type of rational approximations where the expanded function is approximated as a ratio of two polynomial functions $P(x)$ and $Q(x)$ of order $L$ and $M$ (labeled as $[L/M]$ for short), respectively. This type of approximation is suitable for functions/systems containing resonant poles and zeros in their parametric space and it is widely used for approximating a plethora of physical systems~\cite{bender2013advanced}.}

{
The main findings of this Mie--Pad\'e expansion scheme can be illustrated by the following example, where the $[3/3]$ Pad\'e approximant for the first electric dipole Mie coefficient $a_1$ is given in the following form, 
}
\begin{equation}\label{eq:Pade33}
a^{[3/3]}_1=\frac{a^T_1}{1+[y^2]+a^T_1}
\end{equation}
where $a^T_1$ is the static expansion of the Mie coefficient (Eq.~(\ref{eq:dpolar}))~\cite{Tzarouchis2016c} and $[y^2]$ is a second order term presented in details in the next section. {
This type of representation captures both dynamic depolarization and radiative damping effects, exhibiting similarities with other proposed models, see for example in~\cite{Meier1983,deVries1998,Kelly2003,Carminati2006,LeRu2013}. However, in our analysis these dynamic corrections occur by the analysis of the Mie--Pad\'e coefficients, a manifestation of the intrinsic character of these mechanisms to the scattering process. The same Pad\'e approximant scheme can be practically used for the analysis of the scattering amplitudes of arbitrary shaped scatterers possessing a resonant spectrum.}

\subsection{Size-induced dynamic depolarization effects}
{
Before the analysis of the depolarization effects for a core-shell sphere, it is important to make some connecting remarks between the Mie--Taylor results obtained and the proposed Mie--Pad\'e expansion. The lowest non-zero Pad\'e expansion that can be found for the first electric dipole term $a_1$ of a core-shell sphere is of the order $[3/0]$ and it is identical to the Maclaurin expansion of Eq.~(\ref{eq:dpolar}). The next term with a different expression is the $[3/2]$ Pad\'e approximant, viz.,
\begin{equation}\label{eq:pade32}
a^{[3/2]}_1=\frac{a^T_1}{1+[y^2]}
\end{equation}
where the $[y^2]$ term reads
\begin{equation}\label{eq:y2}
[y^2]=\frac{3}{5}\frac{c_1\eta^6+c_2\eta^5+c_3\eta^3+c_4}{c_5\eta^6+c_6\eta^3+c_7}y^2
\end{equation}
where the exact values of coefficients $c_1$ to $c_7$ can be found in Appendix B. 
}
{
Since Eq.~(\ref{eq:y2}) is a complicated expression involving both the thickness and the core permittivity, some physical intuition can be gained by applying the same methodology as above, i.e., expand the $[y^2]$ term for very thick  and thin shells. For the first case the depolarization term reads 
\begin{equation}\label{eq:y2thick}
 [y^2]=-\frac{3}{5}\frac{\varepsilon_2-2}{\varepsilon_2+2}y^2
\end{equation}
being identical to the dynamic depolarization term of the solid sphere~\cite{Tzarouchis2016c}. This result is of no surprise since the thick-shell limit approaches naturally the solid sphere case. Note that this term is independent of the core material $\varepsilon_1$, implying that for thick shells the depolarization terms are mainly affected by the shell material.
}
{
Interestingly, for the thin-shell case we distinguish two different cases. First, a hollow core ($\varepsilon_1=1$) gives  
\begin{equation}
 [y^2]=\frac{1}{5}y^2\frac{4\varepsilon_2+1}{2\varepsilon_2+1}
\end{equation}
while the $[y^2]$ term is 
\begin{equation}
 [y^2]=-\frac{3}{5}\frac{\varepsilon_1-2}{\varepsilon_1+2}y^2
\end{equation}
when core permittivity is $\varepsilon_1$. In the first hollow case the results indicate that the depolarization effects become negligible for shell permittivity values close to $\varepsilon_2=-1/4$ while a resonance is exhibited for values close to $\varepsilon_2=-1/2$. On the other hand, the existence of core material affects the dynamic depolarization is similar to the case of Eq.~(\ref{eq:y2thick}), exposing a zero depolarization values for $\varepsilon_1=2$.
}
{
In a similar manner by analyzing Eq.~(\ref{eq:pade32}) we obtain the following pole conditions for thick ($\eta\rightarrow0$)
\begin{equation}\label{eq:pole_depol1}
 \varepsilon^+=-\frac{1}{2}-\frac{3}{10}y^2\left(1+\frac{7}{10}\eta^3\right)+\frac{3}{2}\eta^3+...
\end{equation}
\begin{equation}
 \varepsilon^-=-2-\frac{12}{5}y^2-6\eta^3
\end{equation}
and for thin shells ($\eta_c\rightarrow0$)
\begin{equation}
 \varepsilon^+=-\frac{2}{3}\eta_c-\frac{4}{15}y^2\eta_c
\end{equation}
\begin{equation}\label{eq:pole_depol4}
 \varepsilon^-=\frac{1}{2}-\frac{3}{2\eta_c}\left(1+\frac{7}{10}y^2\right)-\frac{1}{10}y^2
\end{equation}
indicating that in both cases the symmetric resonance $\varepsilon^-$ exhibits a stronger red-shifting compared to the antisymmetric case.
}

\subsection{Radiative damping}
{
The next interesting Pad\'e expansion is the $[3/3]$ as can be seen in Eq~(\ref{eq:Pade33}), where an extra term of $y^3$ order appears, having the same value of the static polarizability shown in Eq.~(\ref{eq:dpolar}). The radiative damping process is an intrinsic mechanism, required by the conservation of energy, affecting the plasmonic (or any other type) of radiative resonances~\cite{jackson1975electrodynamics} of a scatterer/antenna. 
}

{
Any scatterer can be seen as an open resonator possesing conditions where the scattered fields resonate. The scattering poles of such resonators are generally complex called, often called as \emph{natural frequencies}~\cite{stratton2007electromagnetic}; passivity requires the resonant frequency to be complex. The resonances in our case are studied from the shell permittivity perspective, therefore the appearance of this extra term, i.e, the radiation reaction, will result in an imaginary part of the permittivity condition, even for the lossless case. Note that this term dictates the width of the resonance and the amount of losses required for obtaining maximum absorption in the case of a lossy material~\cite{Tretyakov2014,Tzarouchis2016c,osipov2017modern}.}

{
Turning into the pole condition analysis, by neglecting the $[y^2]$ terms of the first $[3/3]$ Pad\'e approximant we obtain the following pole condition for a thick-shell, hollow ($\varepsilon_1=1$) case:
\begin{equation}\label{eq:rad1}
 \varepsilon^+=-\frac{1}{2}+\frac{3}{2}\eta^3-iy^3\eta^3
\end{equation}
\begin{equation}\label{eq:rad2}
 \varepsilon^-=-2-6\eta^3-2iy^3\left(1+\eta^3\right)
\end{equation}
The complete expression can be restored by including the dynamic depolarization terms, extracted in Eqs.~(\ref{eq:pole_depol1})--(\ref{eq:pole_depol4}). 
}

The dynamic correction term of the symmetric resonance is identical with the solid sphere case~\cite{Tzarouchis2016c}, where the radiative damping term exhibits a $y^3$ dependence including an additional volume ($\eta^3$) thickness contribution. The antisymmetric resonance exhibits a similar size volume dependence, demonstrating a somewhat different dependence, vanishing for very small ratio values.

Similarly, for the thin-shell case we have
\begin{equation}\label{eq:rad3}
 \varepsilon^+=-\frac{2}{3}\eta_c-\frac{2}{9}i\eta_cy^3
\end{equation}

\begin{equation}\label{eq:rad4}
 \varepsilon^-=\frac{1}{2}-\frac{3}{2\eta_c}-i\frac{y^3}{\eta_c}
 \end{equation}
where both imaginary terms have a first order dependency with respect to the ratio. 

All the imaginary terms in Eqs.~(\ref{eq:rad1}),~(\ref{eq:rad2}),~(\ref{eq:rad3}), and~(\ref{eq:rad4}) reveal a very interesting result regarding the radiative damping process: a hollow core-shell structure exhibits an extra degree of freedom for engineering the scattering behavior. This can be used for reverse engineering purposes, e.g., searching for the best suitable material or finding whether the core-shell structures have better absorption performance than the solid ones for a given material. Simply put, the radiative damping term quantifies the amount of the required losses for maximizing the absorption~\cite{Soric2014,Tretyakov2014}, and regulates the maximum linewidth of the scattering process~\cite{Carminati2006,Tzarouchis2017}.

Finally, we will briefly present how the core material influences the radiative damping process. Similar to the static case, as can be found in Section~\ref{sec:core}, the same type of branching is observed here, i.e., at $\varepsilon_1=4$ for thick and $\varepsilon_1=-2$ for thin shells, respectively. For the thick-shell case ($\eta\rightarrow0$) we obtain
\begin{alignat}{2}
\begin{rcases}\label{eq:pole_d}
  \varepsilon^+=-\frac{\varepsilon_1}{2}-\frac{3}{2}\varepsilon_1\frac{\varepsilon_1+2}{\varepsilon_1-4}\eta^3-iy^3g^+(\varepsilon_1,\eta)
  \\
  \\
  \varepsilon^-=-2+6\frac{\varepsilon_1+2}{\varepsilon_1-4}\eta^3-2iy^3g^-(\varepsilon_1,\eta)
\end{rcases}
\text{for}~\varepsilon_1<4
\end{alignat}
with $g^\pm$ being a function of $\varepsilon_1$ and $\eta$, viz., 
\begin{equation}\label{eq:g+}
 g^+(\varepsilon_1,\eta)=9\frac{\varepsilon_1^2}{(\varepsilon_1-4)^2}\eta^3
\end{equation}
\begin{equation}\label{eq:g-}
 g^-(\varepsilon_1,\eta)=1-3\frac{\varepsilon_1(\varepsilon_1+4)-8}{(\varepsilon_1-4)^2}\eta^3
\end{equation}
For a fixed $\eta$ value Eqs.~(\ref{eq:g+}) and~(\ref{eq:g-}) exhibit their maximum and minimum values for $\varepsilon_1\rightarrow0$. 
As a reminder the symmetric damping function Eq.~(\ref{eq:g-}) mutually flips with the antisymmetric function Eq.~(\ref{eq:g+}) when $\varepsilon_1>4$.

In a similar manner the thin-shell case ($\eta_c\rightarrow0$) gives the following pole conditions, 
\begin{alignat}{2}
\begin{rcases}\label{eq:pole_dc}
 \varepsilon^+=-2\frac{\varepsilon_1}{\varepsilon_1+2}\eta_c-2iy^3g^+_c(\varepsilon_1,\eta_c)
 \\
  \varepsilon^-=-\frac{\varepsilon_1}{2}+\frac{\varepsilon_1+2}{2\eta_c}-iy^3g^-_c(\varepsilon_1,\eta_c)
\end{rcases}
~\text{for }~\varepsilon_1>-2
\end{alignat}

with

\begin{equation}\label{eq:gc+}
 g^+_c(\varepsilon_1,\eta_c)=\frac{\varepsilon_1^2}{(\varepsilon_1+2)^2}\eta_c
\end{equation}

\begin{equation}\label{eq:gc-}
 g^-_c(\varepsilon_1,\eta_c)=\frac{1}{\eta_c}-\frac{4}{3}\eta_c\frac{\varepsilon_1(\varepsilon_1-2)-2}{(\varepsilon_1+2)^2}
\end{equation}
As before, Eq.~(\ref{eq:gc+}) has a zero at $\varepsilon_1=0$ while Eq.~(\ref{eq:gc-}) exhibits a minimum value at this point, where the symmetric and antisymmetric resonances mutually exchange their behavior when $\varepsilon_1<-2$.

{
The extracted $g^\pm(\varepsilon_1,\eta)$ functions demonstrate a couple of very interesting peculiarities. First, by adjusting the core
material and ratio we can practically adjust the linewidth and resonant maximum absorption of both symmetric and antisymmetric resonances. This is a particularly attractive feature that enhances the importance of core-shell versus solid scatterers. Moreover, these changes can take place even for very thick shells, having a large impact on the overall performance of the scatterer. Therefore, combining Eqs. (\ref{eq:pole_d})--(\ref{eq:gc-}) one can extract physically intuitive information about the proper design and implementation of small core-shell scatterers. 
}

{
A second point of interest can be derived by comparing the radiative damping and dynamic depolarization term. The first is affected by the core material, even for vanishingly small core sizes, being in stark contrast with the dynamic depolarization mechanism, where for thick shell there are no core material effects. This fact can be potentially used for adjusting the radiative characteristics (linewidth, absorption), preserving at the same time the amount of depolarization effects of a scatterer.
}

\section{Conclusions and Summary}
In this article we presented and discussed the general features regarding the electromagnetic scattering by a small core-shell sphere. Thickness effects for the plasmonic resonances were presented throughout the analysis in the electrostatic (Rayleigh) limit, revealing a series of special characteristics, such as the resonant trends of the symmetric and antisymmetric dipole resonances. These trends are quantified in terms of resonant conditions with respect to the shell permittivity of the scatterer.

A generalization towards higher order multipoles was given, following the same static-limit approximation for the Mie coefficients for small hollow scatterers. {As a rule-of-thumb, thick shells exhibit an $\eta^{2n+1}$ depedence for all n-multipoles, while thin shells exhibit an $\eta_c$ (or $1/\eta_c$ for the antisymmetric case) dependence for all higher multipoles.} 
 
% The perfect agreement between the electrostatic-derived results and the plasmon hybridization model illustrated the common mathematical origins of these theoretical models; the physical intuition of the hybridization model is properly captured by the electrostatic scattering perspective. 

Consequently, we studied the effects of the core material to the overall scattering revealing several aspects about this resonant phenomenon and its intrinsic mechanisms involved. The existence of a critical permittivity value demonstrated a core-induced peculiarity regarding the character of the resonances in the small ratio limit. This can expand the potential usage of plasmonic core-sell structures and their functionalities.

{
The analysis expanded into including also dynamic effects by utilizing the recently introduced Mie--Pad\'e expansion, where the Mie coefficients are expanded as Pad\'e approximants. This perspective delivered physically intuitive information regarding the dynamic effects of small scatterers, such as the dynamic depolarization and the radiation reaction mechanisms. In this way the electrostatic model (Mie--Taylor) has been expanded, revealing promising results about the scattering process, such as the core effects on the dynamic depolarization and the radiative damping dependencies in both radius ratio and core permittivity. 
}

{
We foresee that both the described analysis and the Mie--Pad\'e expansion will inspire further studies regarding the dynamical dependencies of small scatterers. The results can be readily expanded for the study of other, canonical or non canonical scatterers, through the Pad\'e analysis of their scattering amplitudes. In this way new routes can be carved for the precise engineering of their dynamic mechanisms, towards their implementation to applications that span between radio science, applied chemistry, nanotechnology, and beyond.
}

% \begin{table}[h]
% \centering
% \caption{Summary of the main qualitative behavior of the plasmonic resonances as a function of the radius ratio, the core material, and the size of the core-shell structure}
% \begin{tabular}{p{11mm}p{10mm}p{15mm}p{15mm}c} \toprule
%        Dependencies & &  & antisymmetric & symmetric \\ 
%        & & & $\left(\varepsilon^+\right)$ & $\left(\varepsilon^-\right)$ \\ \midrule 
%       ratio $\left(\eta\right)$ & &  $\eta\rightarrow0$ & {red-shift} & \textcolor{blue}{blue-shift}\\
%          & &  & $\left(-\frac{1}{2}\right)$ & $\left(-2\right)$ \\ \cmidrule{3-5}
% 	     & &  $\eta_c\rightarrow0$ & \textcolor{blue}{blue-shift} & {red-shift} \\ 
%         & & & $\left(0^-\right)$ & $\left(-\infty\right)$ \\ \cmidrule{1-5}
% 	core mat. & $\eta\rightarrow0$ & $\varepsilon_1<\varepsilon^{\eta}_\text{br}$ & {red-shift} & $\approx$ const.\\ \cmidrule{3-5}
% $\left(\varepsilon_1\right)$ &  & $\varepsilon_1>\varepsilon^{\eta}_\text{br}$ & $\approx$ const. & {red-shift}\\ \cmidrule{2-5}
% 		  & $\eta_c\rightarrow0$ & $\varepsilon_1<\varepsilon^{\eta_c}_\text{br}$ & $\approx$ const. & {red-shift}\\ \cmidrule{3-5}
% 		  &  & $\varepsilon_1>\varepsilon^{\eta_c}_\text{br}$ & {red-shift} &$\approx$ const.\\ \cmidrule{1-5}
%     
%     
%     size $\left(y\right)$     & $y\uparrow$&  & {red-shift} & {red-shift}\\ \bottomrule
%   \end{tabular}
%   \label{tab:1}
% \end{table}

\section*{Appendix A: Connecting the electrostatic results with the plasmon hybridization model}\label{sec:AppA}
{
The mathematical analysis presented in the above sections emphasized in the thick- and thin-shell dependencies for the permittivity parameter space on its electrostatic limit. A general connection with the electrodynamic scattering and the plasmon hybridization model has been recently presented in~\cite{Thiessen2016}, where the Mie coefficients reorganized, reflecting the plasmon hybridization~\cite{Prodan2003}. 
Here we provide a simple, tutorial-like connection between the aforementioned, static-derived, results and the plasmon hybridization model~\cite{Prodan2003}, where the trends for thin and thick cases are demonstrated. The main emphasis is given to the simplicity and the physical intuition provided by this perspective.
}

{
Assuming that the free (conduction) electrons constitute an incompressible, irrotational, and charged fluid density (plasma)~\cite{Prodan2004}, plasmonic resonances emerge as the oscillatory solutions analogous to the oscillating modes of a harmonic oscillator~\cite{Prodan2004,Mukhopadhyay1975} described by the Laplace equation; its solutions give a set of harmonic functions. The term hybridization implies that the problem is formulated with a Lagrangian method, including both kinetic and potential energy (hybridization) in its system description~\cite{Prodan2004}.
For the case of a spherical hollow core-shell structure, after a significant amount of calculations, these resonant frequencies are described by the following condition~\cite{Mukhopadhyay1975}
\begin{equation}\label{eq:hyb}
 \omega_{n\pm}^2=\frac{\omega^2_B}{2}\left[1\pm\frac{1}{2n+1}\sqrt{1+4n\left(n+1\right)\eta^{2n+1}}\right]
\end{equation}
where $\omega_{n+}^2$ and $\omega_{n-}^2$ is the symmetric and antisymmetric resonances for a given background frequency (plasma frequency) $\omega_B$ and a given multipole, i.e., dipole ($n=1$), quadrupole ($n=2$) and so on~\cite{Roman-Velazquez2011}.}

{
Let us now consider a lossless Drude material dispersion model, i.e., $\varepsilon= 1-\frac{\omega^2_B}{\omega^2}$. By inserting Eq.~(\ref{eq:hyb}) to the material dispersion model we obtain two discrete resonances
\begin{equation}\label{eq:hyb_epsilon1}
 \varepsilon^{\pm}_{n}=1-\frac{2}{1\pm \frac{\sqrt{4 n (n+1) \eta ^{2 n+1}+1}}{2n+1}}
\end{equation}
 described as a function of the radius ratio $\eta$. For the dipole case the following expression is obtained
\begin{equation}\label{eq:hyb_epsilon2}
 \varepsilon^{\pm}_{1}=1-\frac{2}{1\pm \frac{1}{3} \sqrt{1+ 8 \eta ^3}}
\end{equation}
For $\eta\rightarrow0$ the resonances go to $\varepsilon^+_1\rightarrow-\frac{1}{2}$ and $\varepsilon^-_1\rightarrow-2$, respectively, while for $\eta\rightarrow1$ we obtain $\varepsilon^+_1\rightarrow0^-$ and $\varepsilon^-_1\rightarrow-\infty$. 
}

{
Clearly these trends are visible also in the the static model approach described in Section~\ref{sec:static}. A straightforward connection with the static results is obtained with the  expansion of Eq.~(\ref{eq:hyb_epsilon2}) with respect to the radius ratio; for thick shells we obtain
\begin{equation}
 \varepsilon^+_1=-\frac{1}{2}+\frac{3 \eta ^3}{2}
\end{equation}
\begin{equation}
 \varepsilon^-_1=-2-6 \eta ^3
\end{equation}
while for thin shells ($\eta_c$) we have
\begin{equation}
 \varepsilon^+_1=-\frac{2}{3} \eta_c
\end{equation}
\begin{equation}
 \varepsilon^-_1=\frac{1}{2}-\frac{3}{2\eta_c}-\frac{1}{3}\eta_c
\end{equation}
corresponding exactly to the static results of Eqs. (\ref{eq:static_thick1}), (\ref{eq:static_thick2}), (\ref{eq:static_thin1}), and (\ref{eq:static_thin2}), respectively.  This fact can be also seen by the comparison between Eq.~(\ref{eq:static}) and Eq.~(\ref{eq:hyb_epsilon1}), where both expressions are mathematically equivalent. Note that all the higher order terms were truncated. 
}

{
It is obvious that the electrostatic analysis captures the essence of the hybridization model, since both models exhibit spherical harmonic solutions (Laplace equation)~\cite{Prodan2003,Sihvola1988}. As a consequence, the electrodynamic perspective (Mie theory) is practically equivalent with both models in the small-size limit. For completeness, we derive the frequency (energy) equivalent relations for both thick and thin cases, viz., 
\begin{alignat}{2}
\begin{rcases}
\omega^2_{1+}=\frac{2}{3}\omega^2_B(1+\eta^3)
\\
\omega^2_{1-}=\frac{1}{3}\omega^2_B(1-2\eta^3)
\end{rcases} ~\text{for }~\eta\rightarrow0
\end{alignat}
and
\begin{alignat}{2}
\begin{rcases}
 \omega^2_{1+}=\omega^2_B(1-\frac{2}{3}\eta_c)
 \\
  \omega^2_{1-}=\omega^2_B\frac{2}{3}\eta_c
\end{rcases}
~\text{for }~\eta_c\rightarrow0
\end{alignat}
}
\section*{Appendix B: dynamic depolarization terms}\label{sec:AppB}
{
In this section the coefficients $c_1$ to $c_7$ used in Eq.~(\ref{eq:y2}) are presented. These coefficients are functions of both core and shell permittivities $\varepsilon_1$ and $\varepsilon_2$, 
\begin{equation}
 c_1=2 d_1d_2^2 \left(4\varepsilon_2+1\right)
\end{equation}
\begin{equation}
 c_2=-9d_2\varepsilon_2^2\left(\varepsilon_1-2\varepsilon_2\right)
\end{equation}
\begin{equation}
 c_3=-d_1d_2d_3\left(7\varepsilon_2+4\right)
\end{equation}
\begin{equation}
 c_4=-d_1d_3^2 \left(\varepsilon_2-2\right) 
\end{equation}
\begin{equation}
 c_5=2 d_1d_2^2 \left(2\varepsilon_2+1\right)
\end{equation}
\begin{equation}
 c_6=d_2d_3\left(4\varepsilon_2^2+\varepsilon_2+4\right)
\end{equation}
\begin{equation}
 c_7=d_1d_3^2 \left(\varepsilon_2+2\right) 
\end{equation}
with 
\begin{equation}
 d_1=\varepsilon_2-1\text{, }d_2=\varepsilon_1-\varepsilon_2\text{, and } d_3=\varepsilon_1+2\varepsilon_2
\end{equation}
}

\balance



\bibliographystyle{IEEEtran}
\bibliography{IEEEabrv,Notes_Pade}

%\begin{thebibliography}{1}


%\end{thebibliography}

 \begin{IEEEbiographynophoto}{Dimitrios C. Tzarouchis} 
%  ('S-13) received the
% Diploma (M.Sc.) degree in electrical and computer engineering from the Aristotle University of
% Thessaloniki, Greece, in 2013. Since 2015 he is a doctoral candidate with the Depatment of Electronics and Nanoengineering, Aalto University, Finland. His main research interests include electromagnetic scattering theory, plasmonics, and antenna theory. He is recipient of the 2015 IEEE Antennas and Propagation Society Doctoral Research Grant and the Best Student Paper Award (2nd price) in PIERS 2017 in St. Petersburg, Russia. 

\end{IEEEbiographynophoto}
 \begin{IEEEbiographynophoto}{Ari Sihvola} 
%  ('F-06) received the degree of Doctor of Technology in 1987 from the Helsinki University
% of Technology, Finland (presently Aalto University). Besides working for TKK, Aalto, and the
% Academy of Finland, he was visiting engineer in the Research Laboratory of Electronics of the
% Massachusetts Institute of Technology, Cambridge, in 1985–1986. In 1990-1991, he worked as
% a visiting scientist at the Pennsylvania State University, State College. In 1996, he was visiting
% scientist at the Lund University, Sweden. He was visiting professor at the Electromagnetics and
% Acoustics Laboratory of the Swiss Federal Institute of Technology, Lausanne (academic year
% 2000-01), in the University of Paris 11, in Orsay (June 2008), and in the Unviersity of Rome La
% Sapienza (May-June 2015). His research interests include waves and fields in electromagnetics,
% modeling of complex media an metamaterials, remote sensing, and radar applications. He is
% presently professor in the School of Electrical Engineering at the Aalto University, Finland.
 \end{IEEEbiographynophoto}


%\vfill


\end{document}